\documentclass[pdflatex,sn-mathphys-num,iicol]{sn-jnl}

\usepackage{graphicx}%
\usepackage{multirow}%
\usepackage{amsmath,amssymb,amsfonts}%
\usepackage[title]{appendix}%
\usepackage{xcolor}%
\usepackage{textcomp}%
\usepackage{booktabs}%
\usepackage{array}%
\usepackage{tikz}%
\usetikzlibrary{arrows.meta,positioning,fit,calc}

\graphicspath{{figures/}}
\newcommand{\dg}{\ensuremath{^{\circ}}}
\newcommand{\vx}{\mathbf{x}}
\newcommand{\vc}{\mathbf{c}}
\newcommand{\dd}{\ensuremath{|\Delta\theta|}}
\newcommand{\Co}{\ensuremath{^{60}\mathrm{Co}}}

\begin{document}

\title[Label-free self-calibration of an HPGe tracking array]{Label-free self-calibration of a segmented HPGe $\gamma$-ray tracking array from the Compton scattering constraint}

\author*[1,2,3]{\fnm{Mojahed} \sur{Abushawish}}\email{mojaheda@ucm.es}
\author[3]{\fnm{J\'er\'emie} \sur{Dudouet}}
\author[3]{\fnm{Olivier} \sur{St\'ezowski}}
\author[1,2]{\fnm{Jos\'e Manuel} \sur{Ud\'ias}}

\affil[1]{\orgdiv{Grupo de F\'isica Nuclear, EMFTEL \& IPARCOS}, \orgname{Universidad Complutense de Madrid}, \orgaddress{\street{CEI Moncloa}, \city{Madrid}, \postcode{28040}, \country{Spain}}}

\affil[2]{\orgname{Health Research Institute of the Hospital Cl\'inico San Carlos (IdISSC)}, \orgaddress{\city{Madrid}, \postcode{28040}, \country{Spain}}}

\affil[3]{\orgdiv{IP2I Lyon, UMR 5822}, \orgname{Universit\'e Claude Bernard Lyon 1, CNRS/IN2P3}, \orgaddress{\city{Villeurbanne}, \postcode{69100}, \country{France}}}

\abstract{\unboldmath Pulse-shape analysis in segmented HPGe $\gamma$-ray tracking arrays such as AGATA relies on reference signals for known interaction positions, which are costly to measure and only approximate when calculated. In a simulation of 51 AGATA crystals, we show that the interaction positions can be learned from the pulses without position labels. For a \Co\ photon that is Compton scattered in one segment and absorbed in another, the scattering angle follows both from the deposited energies and from the positions of the source and of the two interactions. A neural network shared by all crystals predicts one position per fired segment and is trained only to make these two angles agree; the unmeasured order of the two interactions and the selection of the pairs are revised during training. The pulses, synthesised from AGATAGeFEM bases for $2.79\times10^{6}$ pairs of fired segments whose energies add up to a \Co\ line, selected from $4\times10^{7}$ simulated decays, include noise, cross-talk, trigger-time jitter and differences between crystals. For the hits of held-out pairs, the mean three-dimensional position error falls from 10.4~mm with the segment centres to 1.65~mm, with a FWHM of 1.0--1.3~mm in each coordinate, compared with 1.07~mm for the same network trained on the simulated positions. The constraint determines the positions along the crystal axes only weakly, and at small scattering angles it pulls low-energy first interactions towards the source, which is why deposits below 200~keV are excluded. Even with the simulated positions, the two angles differ, mainly because a segment with several interactions is described by a single position and energy rather than because of Doppler broadening.}


\maketitle

\section{Introduction}\label{sec:intro}

$\gamma$-ray tracking arrays of highly segmented high-purity germanium (HPGe) detectors, such as AGATA~\cite{Akkoyun2012,Clement2017} and GRETINA~\cite{Lee2003,Paschalis2013}, reconstruct the path of each photon from the energies and positions of its interactions~\cite{LopezMartens2004,Korichi2019}. The positions are obtained by pulse-shape analysis (PSA), which compares the signals of the core and of the 36 segments of a crystal with a \emph{basis} of reference signals for known positions~\cite{Venturelli2004,Boston2023}. In-beam measurements give a position resolution of about 5~mm full width at half maximum (FWHM) at energies around 1~MeV~\cite{Recchia2009,Soderstrom2011}.

The basis is either calculated or measured, and both have limitations. Calculated bases~\mbox{\cite{Bruyneel2016,Ljungvall2021,Schlarb2011JASS}} depend on the impurity profile, the charge-carrier mobilities~\cite{Mihailescu2000,Bruyneel2006a} and the electronics, including cross-talk~\cite{Bruyneel2009xt}, which are known only approximately for an individual crystal and affect the PSA resolution~\cite{Lewandowski2019}. Measured bases from scanning tables~\cite{Nelson2007,Ha2013,Crespi2008,Goel2011,HernandezPrieto2016,DeCanditiis2021} require a dedicated measurement of each crystal before it is mounted and cannot follow later changes such as neutron damage~\cite{Descovich2005a,Bruyneel2013}. Methods based on the distribution of measured pulses~\cite{Desesquelles2011,Desesquelles2013,Li2018} avoid scanning but rely on an assumed spatial distribution of the interactions.

A tracking array offers an additional, physical constraint. For a photon that is Compton scattered and then absorbed, the scattering angle follows both from the deposited energies and from the positions of the source and of the two interactions~\cite{Compton1923}. This redundancy allows the positions to be calibrated without knowing them in advance, a \emph{self-calibration}. Heil \emph{et al.}\ proposed it for groups of interactions with similar pulses and demonstrated it for a segmented spherical shell, with the groups formed from the simulated positions~\cite{Heil2018}. Chen \emph{et al.}\ developed it for AGATA by clustering the pulses, optimising the position of each cluster with tracked events and using the result as a new basis for the PSA~\cite{Chen2025}. On their simulated pulses, the PSA reaches about 2~mm FWHM with the basis from which the pulses were generated, and with the self-calibrated basis about 3~mm FWHM without noise and about 3.5~mm with a noise of 3~keV RMS, the level also used here. These PSA resolutions are not directly comparable with our position errors (Sect.~\ref{sec:comparison}).

Neural networks, increasingly used in nuclear physics~\cite{Boehnlein2022}, can represent the mapping from pulses to positions as a continuous function. In previous work, networks trained with scanning data, which give only two of the three coordinates of each interaction, characterised AGATA crystals with an accuracy of about 2~mm~\cite{Abushawish2026}. Here we remove the position labels altogether: the training signal is the Compton relation itself, in the spirit of physics-informed learning~\cite{Karniadakis2021}. A similar approach, PhysiCal-PET, calibrates the depth of interaction in a positron emission tomography scanner without depth labels~\cite{Abushawish2026PET}. A single network is shared by all segments and crystals. Unlike in Ref.~\cite{Chen2025}, the pulses are not clustered and the network provides the positions directly rather than a new basis. The order of the two interactions, which is not measured, and the selection of the pairs are revised during training with the reconstructed positions.

We assess the method with simulated data for 51 AGATA crystals irradiated by a single \Co\ source, with a detector response modelled after Ref.~\cite{Chen2025}. The self-calibration does not use the simulated positions, but they allow its accuracy to be measured directly. Sections~\ref{sec:simulation}--\ref{sec:results} describe the simulated data, the method and the results, including a comparison with a supervised network trained on the simulated positions, and Sect.~\ref{sec:discussion} discusses what the constraint determines and the limitations of the method.

\begin{figure*}[t]
\centering
\includegraphics[width=\textwidth]{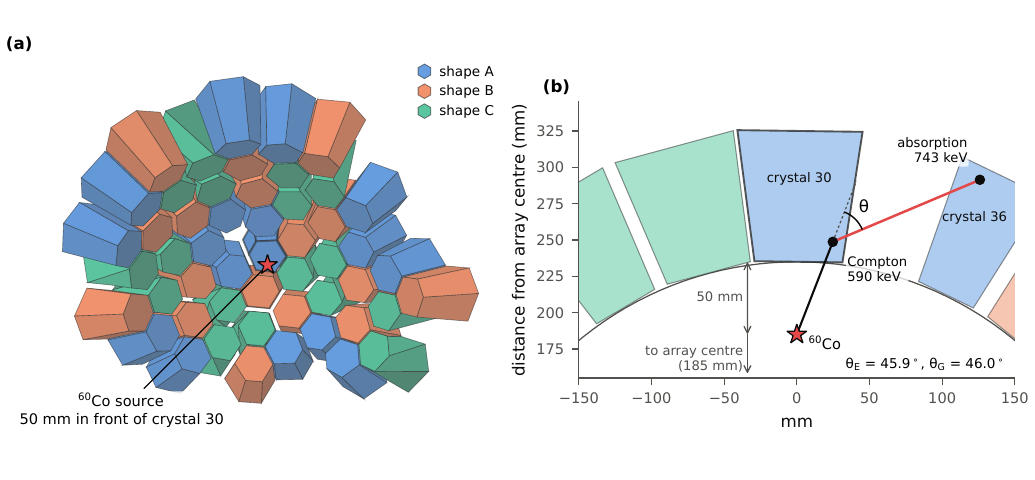}
\caption{Simulated set-up. (a) The 51 crystals, coloured by shape, viewed from beyond the array centre towards their front faces. (b) Section to scale through the array centre and the source, cutting crystals 30 and 36, with an example pair of hits from the initial sample (Sect.~\ref{sec:selection}). The arc marks the front faces. A 1332~keV photon deposits 590~keV by Compton scattering in crystal 30 and 743~keV by absorption in crystal 36. It has one interaction in each crystal, both within 2~mm of the section plane. Black: incident direction; red: scattered direction. The angles $\theta_E$ from the two energies, Eq.~\eqref{eq:compton}, and $\theta_G$ from the source and the two simulated positions, Eq.~\eqref{eq:geometry}, agree within $0.1\dg$ (values at the bottom right)}
\label{fig:geometry}
\end{figure*}
\section{Simulated data}\label{sec:simulation}

\subsection{Array, source and physics}\label{sec:array}

The simulated array consists of 17 AGATA triple clusters at their positions in the $4\pi$ configuration~\cite{Akkoyun2012,Farnea2010}, i.e.\ 51 crystals, 17 of each of the asymmetric shapes A, B and C, with their front faces 235~mm from the array centre (Fig.~\ref{fig:geometry}). Each crystal is a tapered hexagonal coaxial detector, 90~mm long, with 36 segments: six azimuthal sectors A--F (unrelated to the shape letters) in each of six rings, numbered 1--6 from the front, so that a segment is named, for example, B4. The inactive germanium, the air gap and the 0.8~mm aluminium capsule are included; the cryostats and support structures are not.

A point-like \Co\ source is placed 50~mm in front of crystal 30, of shape A, on the line from the centre of its front face to the array centre. The front faces of the other 50 crystals are 73--241~mm from the source, and since their axes point at the array centre rather than at the source, the photons enter them obliquely. Each decay emits the 1173.24 and 1332.51~keV photons isotropically and without angular correlation. An \emph{event} is one decay, with the interactions of both photons; $4\times10^{7}$ events are simulated with SToGS~\cite{SToGS} and Geant4 11.2~\cite{Agostinelli2003,Allison2006,Allison2016}. Electromagnetic processes follow the option-4 physics constructor with a production cut of 100~\textmu m. Compton scattering is described by the Monash model, which includes Doppler broadening~\cite{Brown2014,Ribberfors1975}, and photoabsorption and Rayleigh scattering by the Livermore models. Atomic de-excitation (fluorescence X-rays and Auger electrons) is switched off.

\subsection{Interaction recording and segment packing}\label{sec:packing}

An \emph{interaction} is recorded whenever a photon of the decay undergoes Compton scattering, photoabsorption or pair production in the active germanium. Its position is the end point of the photon step, and its energy is the energy deposited in the step plus the kinetic energy of the electrons and positrons created in it, whose tracks are thus treated as points. Rayleigh scattering, which deposits no energy, and scattering in passive material deflect a photon without being recorded; such deflections can make the angle from the source and the recorded positions differ from the angle from the energies, even though these positions are exact (Sect.~\ref{sec:floor}).

\begin{figure*}[t]
\centering
\includegraphics[width=\textwidth]{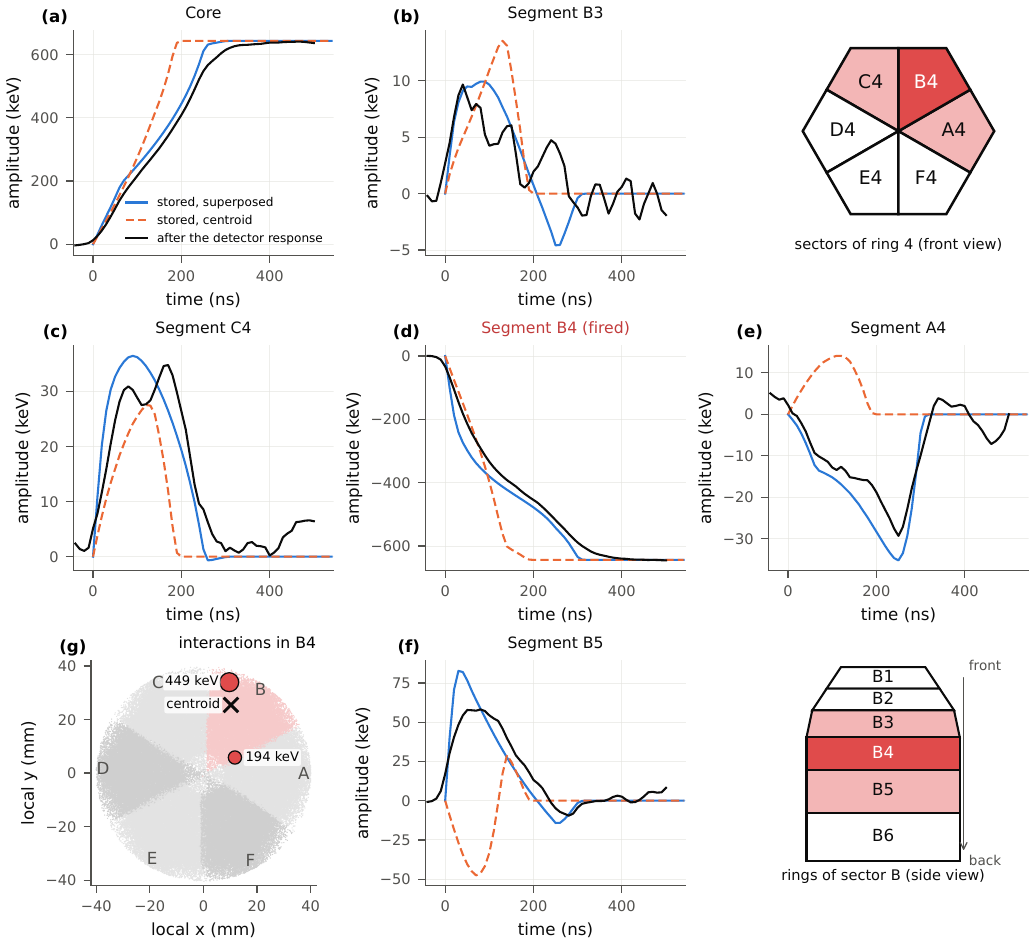}
\caption{Pulse view of a 643~keV hit in segment B4 of a crystal of shape C, whose partner hit lies in another crystal. The hit contains two interactions: 449~keV at 35~mm from the crystal axis and 194~keV at 13~mm. (a) Core; (d) the fired segment B4; (b), (c), (e), (f) its four first neighbours, placed as in the diagrams on the right; (g) the two interactions (circles) and their packed position (cross, labelled ``centroid''), in the transverse plane. Curves: superposed pulse (blue) and centroid pulse (orange, dashed), both as saved (``stored'' in the legend); superposed pulse after the detector response (Sect.~\ref{sec:noise}), aligned on the estimated trigger time, before the final normalisation (black). The centroid pulse rises faster than the superposed pulse in (a) and (d), and its transients in (e) and (f) have the opposite sign}
\label{fig:pulses}
\end{figure*}

Each interaction is assigned to a segment by locating its position in the 1~mm grid of the pulse basis of its crystal shape (Sect.~\ref{sec:pulses}). All interactions of an event in the same segment are packed into one \emph{hit}, with energy and position
\begin{equation}
 E_{\rm seg}=\sum_j\varepsilon_j,\qquad
 \overline{\vx}_{\rm seg}=\frac{1}{E_{\rm seg}}\sum_j\varepsilon_j\vx_j ,
 \label{eq:packing}
\end{equation}
where $\varepsilon_j$ and $\vx_j$ are the energies and positions of its interactions. A segment with a hit is a \emph{fired segment}, and a hit with more than one interaction is a \emph{multi-interaction hit}. The energy-weighted centroid $\overline{\vx}_{\rm seg}$, the \emph{packed position}, is the true (simulated) position of the hit; for a multi-interaction hit it in general coincides with none of the interactions (Fig.~\ref{fig:pulses}g). The \emph{position error} of a hit is the three-dimensional distance between its reconstructed and its packed position. After packing, the energy of each hit is smeared with a Gaussian whose FWHM, typical of AGATA segments, rises linearly from 1.2~keV at 60~keV to 2.1~keV at 1333~keV; the positions are not smeared.

\subsection{Event and pair selection}\label{sec:selection}

\begin{table*}[t]
\caption{Selection of the pairs, step by step. Counts are numbers of events in the first three rows and numbers of pairs below. ``Kept'' is the percentage of the row above, or of the base sample for step~7. The step-7 counts include the validation pairs. Of the $3.53\times10^{6}$ stored pairs, $1.95\times10^{6}$ lie on the 1173~keV line and $1.58\times10^{6}$ on the 1332~keV line}
\label{tab:cutflow}
\centering\small
\begin{tabular}{@{}l>{\raggedright\arraybackslash}p{0.36\textwidth}>{\raggedright\arraybackslash}p{0.25\textwidth}rr@{}}
\toprule
Step & Selection & Information used & Count & Kept (\%)\\
\midrule
\multicolumn{5}{@{}l}{\emph{Event filter (steps~1--3), applied to the simulated events}}\\
 & Simulated decays (events) & & $4\times10^{7}$ & \\
 & Events with at least one interaction in the germanium & & $3.17\times10^{7}$ & 79.3\\
 & Events with at least two fired segments & fired segments & $2.57\times10^{7}$ & 81.1\\
 & Candidate pairs: all pairs of hits in these events & & $1.57\times10^{8}$ & \\
1 & Photopeak: sum of the two energies within $\pm5$~keV of a \Co\ line & segment energies & $8.84\times10^{6}$ & 5.6\\
2 & Distance: segment centres at most 160~mm apart & segment identifiers, array geometry & $8.13\times10^{6}$ & 91.9\\
3 & No first neighbours in the same crystal & segment identifiers & $4.67\times10^{6}$ & 57.4\\
 & \quad and no second neighbours in the same crystal & segment identifiers & $3.53\times10^{6}$ & 75.6\\
 & Order of the two hits assigned (stored order), no pair removed; result: \emph{stored pairs} & segment energies, segment centres, source position & $3.53\times10^{6}$ & 100\\
\midrule
\multicolumn{5}{@{}l}{\emph{Cuts on the stored pairs (steps~4--6)}}\\
4 & Physical Compton angle: $|\cos\theta_E|\le1.01$ & segment energies & $3.53\times10^{6}$ & 99.95\\
5 & Energy threshold: both hits at least 200~keV; result: \emph{base sample} & segment energies & $2.79\times10^{6}$ & 79.1\\
6 & Segment-centre residual $\dd\le25\dg$; result: \emph{initial sample} & segment energies, segment centres, source position & $2.21\times10^{6}$ & 79.1\\
\midrule
\multicolumn{5}{@{}l}{\emph{Selection during training (step~7), applied to the base sample}}\\
7 & Residual $\dd\le5\dg$ with the network of stage~1, before stage~2 & segment energies, reconstructed positions, source position & $1.41\times10^{6}$ & 50.4\\
 & \quad with the network of stage~2, before stage~3, renewed order & same as above & $1.45\times10^{6}$ & 52.1\\
 & \quad with the network of stage~3, before stage~4, renewed order & same as above & $1.46\times10^{6}$ & 52.4\\
\bottomrule
\end{tabular}
\end{table*}

The calibration uses pairs of hits, a \emph{first hit} with energy $E_1$ and a \emph{second hit} with energy $E_2$, which are assumed to be the first interaction of a photon, a Compton scattering, and the full absorption of the scattered photon. We call these two assumptions the \emph{two-interaction hypothesis}. For pairs that fulfil it, the scattering angle $\theta_E$ derived from the energies agrees with the angle $\theta_G$ given by the source and the two hit positions (Sect.~\ref{sec:constraint}); their difference is the residual ${\Delta\theta=\theta_G-\theta_E}$, and $\dd$ is its absolute value. The pairs are selected in seven steps, summarised in Table~\ref{tab:cutflow}: an event filter applied to the simulated events (steps 1--3), cuts on the pairs stored by this filter (steps 4--6) and a selection renewed during training (step 7). Every step uses only information that is also available in a measurement (column ``Information used'' of Table~\ref{tab:cutflow}); the simulated positions are used only to evaluate the result.

In each event with at least two fired segments, every combination of two hits is a candidate pair, $1.57\times10^{8}$ in total. The event filter keeps a candidate pair if (1) its two energies add up to within $\pm5$~keV of one of the two \Co\ lines, as expected when the photon deposits its full energy in the two segments; (2) the two segment centres are at most 160~mm apart; and (3) the two segments, if they belong to the same crystal, are at least three steps apart, counting the sector difference around the crystal plus the ring difference. Condition~1 removes 94\% of the candidate pairs. Condition~2 rejects distant pairs because they violate the two-interaction hypothesis more often: with the packed positions, used here only to justify the cut, $\dd$ exceeds $10\dg$ for 28--35\% of the base-sample pairs whose segment centres are 20--100~mm apart, but for 57\% of those 140--160~mm apart. Condition~3 rejects first neighbours (one step) and second neighbours (two steps). Since the network input of a hit contains the signals of its first neighbours (Sect.~\ref{sec:pulses}), a partner hit in a first neighbour would add its net charge to one of these channels, and one in a second neighbour a large transient. The filter stores $3.53\times10^{6}$ pairs, the \emph{stored pairs}.

The order of the two hits, which $\theta_E$ requires, is not measured. The filter assigns an order without removing any pair: for each of the two orders it computes $\theta_G$ with the segment centres and compares it with the angle of Eq.~\eqref{eq:compton} evaluated with the energy of the closest \Co\ line, and it keeps the order with the smaller difference, the \emph{stored order}. An order is allowed only if $E_2$ is at least the energy of a photon backscattered at $180\dg$, 210~keV on the 1173~keV line and 214~keV on the 1332~keV line, because a scattered photon cannot have less energy. A pair with a deposit below this energy therefore has only one allowed order.

Of the stored pairs, 22\% are \emph{same-crystal pairs}, whose two hits lie in one crystal and share its core trace. The filter does not require exactly two fired segments, and 61\% of the stored pairs come from events with three or more. The additional segments, which may belong to either photon, are not included in the simulated pulses of the pair (Sect.~\ref{sec:pulses}), whereas in a measurement an additional segment in the crystal of a hit would add to its core and neighbour signals (Sect.~\ref{sec:limitations}). If an additional segment belongs to the photon of the pair, the two hits miss part of its energy, and their sum can fall on the other line, e.g.\ on 1173~keV for a 1332~keV photon (\emph{wrong-line pairs}, Sect.~\ref{sec:truth}).

Three cuts are then applied to the stored pairs in the stored order. (4) Equation~\eqref{eq:compton} must give $|\cos\theta_E|\le1.01$, a tolerance for the energy resolution; this removes 1639 pairs. (5) Both hits must deposit at least 200~keV. The $2.79\times10^{6}$ remaining pairs form the \emph{base sample}, from which all training and evaluation pairs are taken, except for a preliminary training and its validation pairs (Sects.~\ref{sec:training} and~\ref{sec:lowenergy}). Since 200~keV is below both backscatter energies, a hit below 200~keV is always a first hit, and the cut removes only pairs with $\theta_E$ below $21\dg$ on the 1332~keV line or below $24\dg$ on the 1173~keV line. It was introduced because, in the preliminary training, the constraint pulled first hits below 100~keV towards the source (Sect.~\ref{sec:lowenergy}). (6) Before training, the only positions available are the segment centres. With them, $\dd$ has a median of $10.9\dg$ in the base sample, and the $2.21\times10^{6}$ pairs with $\dd\le25\dg$ form the \emph{initial sample}, from which the validation pairs and the first training pool are drawn (Sect.~\ref{sec:training}).

During training, (7) the cut of step 6 is repeated with the reconstructed positions and a limit of $5\dg$: before each of training stages 2, 3 and 4, $\dd$ of every pair of the base sample is recomputed with the network of the previous stage, and the pairs with $\dd\le5\dg$, $1.41$--$1.46\times10^{6}$ or about half of the base sample, form the training pool of the stage. Because the reconstructed positions are much more precise than the segment centres, the tighter limit rejects more pairs that violate the two-interaction hypothesis, and pairs rejected in step 6 can be readmitted. Section~\ref{sec:truth} shows how steps 6 and 7 act on the different kinds of pairs.

Since no step uses the simulated positions, the same selection can be applied to a measurement with a \Co\ source. Three effects absent from the simulation must then be considered: random coincidences between decays add accidental pairs, which step 1 suppresses but does not remove; the segment energies must be corrected for cross-talk before they are summed; and segment thresholds remove small deposits.

\subsection{Pulse synthesis}\label{sec:pulses}

The signals of each hit are interpolated from pulse bases of the three crystal shapes calculated with \mbox{AGATAGeFEM}~\cite{Ljungvall2021}. For charge released at each point of a 1~mm grid, they give the signals of the core and the 36 segments, without preamplifier response, cross-talk or noise, sampled every 10~ns over 540~ns from the interaction time. The \emph{basis pulse} at any position is the inverse-square-distance average of the signals at the eight nearest grid points of the same segment. For each hit, the network receives a \emph{pulse view} of six channels of 55 samples: the core, the fired segment and its four first neighbours, two in the same sector and two in the same ring (one channel is empty in the first and last rings). The segment signals are inverted, so that the core signal is positive and the fired-segment signal negative (Fig.~\ref{fig:pulses}).

Two sets of pulse views are synthesised for every stored pair. \emph{Superposed pulses}, used for training and, unless stated otherwise, for evaluation, are the energy-weighted sums of the basis pulses of all interactions in the two segments of the pair; interactions in other segments of the event are not included. In a same-crystal pair, each view therefore also contains the signals of the partner hit, and the two hits share one core trace. \emph{Centroid pulses} contain a single basis pulse at the packed position of each hit and no partner signal. For a multi-interaction hit, the superposed pulse combines several drift times in the rise of the core and fired-segment signals and transients from several positions in the neighbours, which a single basis pulse at the centroid does not reproduce (Fig.~\ref{fig:pulses}). Centroid pulses are used only for evaluation, to measure how much these two features add to the position error (Sect.~\ref{sec:accuracy}).

\subsection{Detector response}\label{sec:noise}

The detector response turns the stored pulse views into realistic signals, following Ref.~\cite{Chen2025} and the conventions of the AGATA PSA~\cite{Boston2023}. Each view, normalised to the maximum of its core trace, is first multiplied by the energy of its hit (for superposed pulses of a same-crystal pair, by the summed energy of both hits), so that the following effects act on amplitudes in keV:
\begin{enumerate}
\item a stretch of the time axis by a factor drawn per crystal, with mean 1 and standard deviation 2\%, a simplified emulation of crystal-to-crystal differences in the charge drift;
\item a single-pole preamplifier response with a time constant of 35~ns;
\item cross-talk on each neighbour channel: a proportional term, equal to a coefficient times the mean of the fired-segment and core signals, with coefficients around ${-1\times10^{-3}}$~\cite{Bruyneel2009xt} and relative spreads of 30\% between crystals and 10\% between segments, and a differential term, the time derivative of the fired-segment signal times the same coefficient and a factor of $-10$ for a neighbour in the same sector on the same triple preamplifier (rings 1--3 or 4--6), $-1$ for one on the other triple preamplifier and $+1$ for the neighbours in the same ring (the AGATA PSA software multiplies these factors by a further 10 by default, which we omit so that this term stays small compared with the noise);
\item channel gains and delays, drawn with standard deviations of 0.5\% and 2~ns;
\item Gaussian noise of 3~keV RMS per sample, band-limited by the preamplifier response;
\item a three-point smoothing with weights 0.1, 0.8 and 0.1, applied twice, as for the AGATA PSA basis;
\item a trigger time from a straight-line fit to the leading edge of the noisy core signal, on which all channels are re-aligned with a pre-trigger of 40~ns.
\end{enumerate}
The crystal and channel parameters are drawn once and kept fixed, whereas new noise is drawn each time a pulse view is loaded, so that the trigger time jitters from one realisation to the next. For hits whose partner lies in another crystal, the robust standard deviation of this jitter is about 9~ns at 200--300~keV and about 2~ns above 800~keV. Finally, the core trace is rescaled to the amplitude of the fired segment and each view is divided by its largest absolute value.

\section{Compton-constrained learning}\label{sec:method}

\subsection{Angular constraint}\label{sec:constraint}

Under the two-interaction hypothesis, the photon has the energy $E_1+E_2$ before the scattering at the first hit and $E_2$ after it, and the free-electron Compton relation gives the scattering angle
\begin{equation}
 \cos\theta_E=1-m_ec^2\left(\frac{1}{E_2}-\frac{1}{E_1+E_2}\right),
 \label{eq:compton}
\end{equation}
with $m_ec^2=510.999$~keV. The measured sum is used instead of the nominal line energy to keep the relation consistent with the energy resolution of the pair. The geometrical angle follows from the source position $\vx_s$ and the hit positions $\vx_1$ and $\vx_2$ in the global frame,
\begin{equation}
 \theta_G=\operatorname{atan2}\left(\|\mathbf u\times\mathbf v\|,\;\mathbf u\cdot\mathbf v\right),
 \label{eq:geometry}
\end{equation}
with the incident direction $\mathbf u=\vx_1-\vx_s$ and the scattered direction $\mathbf v=\vx_2-\vx_1$. Equation~\eqref{eq:geometry} is evaluated with the segment centres for the stored order and step 6, with the reconstructed positions for the loss, step 7 and the renewed order (Sect.~\ref{sec:training}), and with the packed positions for evaluation. Even with the packed positions, $\dd$ is not zero; we call it then the \emph{true-position residual}. It arises mainly because a multi-interaction hit is described by a single position and energy, and less from Doppler broadening (Sect.~\ref{sec:floor}).

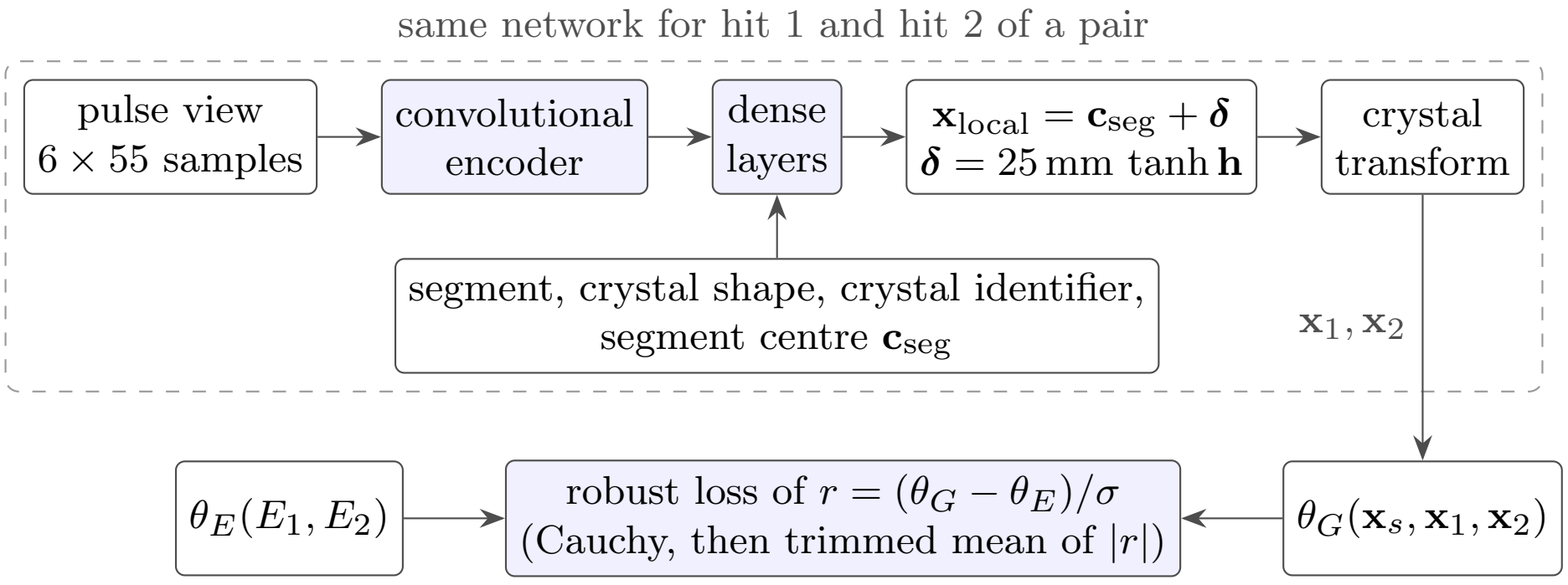
\begin{figure*}[t]
\centering
\begin{tikzpicture}[
  font=\small,
  box/.style={draw=black!70, rounded corners=2pt, align=center, inner sep=3pt, minimum height=9mm, fill=white},
  hl/.style={box, fill=blue!6},
  arr/.style={-{Stealth[length=2mm]}, black!70}]
\node[box] (pulse) {pulse view\\ $6\times55$ samples};
\node[hl, right=5mm of pulse] (enc) {convolutional\\ encoder};
\node[hl, right=5mm of enc] (mlp) {dense\\ layers};
\node[box, right=5mm of mlp] (off) {$\vx_{\rm local}=\vc_{\rm seg}+\boldsymbol\delta$\\ $\boldsymbol\delta=25\,\mathrm{mm}\,\tanh\mathbf h$};
\node[box, right=5mm of off] (glob) {crystal\\ transform};
\node[box, below=5mm of mlp] (ids) {segment, crystal shape, crystal identifier,\\ segment centre $\vc_{\rm seg}$};
\node[draw=black!40, dashed, rounded corners=3pt, fit=(pulse)(enc)(mlp)(off)(glob)(ids), inner sep=4pt,
      label={[font=\small, black!70]above:same network for hit 1 and hit 2 of a pair}] (net) {};
\node[box, below=21mm of glob] (thg) {$\theta_G(\vx_s,\vx_1,\vx_2)$};
\node[hl, left=8mm of thg] (loss) {robust loss of $r=(\theta_G-\theta_E)/\sigma$\\ (Cauchy, then trimmed mean of $|r|$)};
\node[box, left=8mm of loss] (the) {$\theta_E(E_1,E_2)$};
\draw[arr] (pulse) -- (enc);
\draw[arr] (enc) -- (mlp);
\draw[arr] (ids) -- (mlp);
\draw[arr] (mlp) -- (off);
\draw[arr] (off) -- (glob);
\draw[arr] (glob) -- node[left]{$\vx_1,\vx_2$} (thg);
\draw[arr] (thg) -- (loss);
\draw[arr] (the) -- (loss);
\end{tikzpicture}
\caption{Self-calibration scheme. The network (dashed box) predicts the displacement $\boldsymbol\delta$ of each hit from its segment centre. The angle $\theta_G$ from the reconstructed positions is compared with $\theta_E$ through a robust loss of the normalised residual $r$ (Sect.~\ref{sec:loss}). The learned-uncertainty output is not shown. No position labels are used}
\label{fig:method}
\end{figure*}

\subsection{Position network}\label{sec:network}

A single network with $7.12\times10^{5}$ trainable parameters serves all 51 crystals and reconstructs the two hits of a pair independently (Fig.~\ref{fig:method}; settings in Table~\ref{tab:parameters}). A one-dimensional convolutional encoder extracts features from the pulse view of the hit. These are combined with the segment centre $\vc_{\rm seg}$ in the local frame of the crystal, whose $z$ axis is the crystal axis, and with learned embeddings (short trainable vectors) of the segment number, crystal shape and crystal identifier. Dense layers then predict the displacement $\boldsymbol\delta$ of the hit from its segment centre, bounded in each coordinate,
\begin{equation}
 \vx_{\rm local}=\vc_{\rm seg}+\boldsymbol\delta,\qquad \boldsymbol\delta=25~\mathrm{mm}\,\tanh(\mathbf h),
 \label{eq:offset}
\end{equation}
which the known position and orientation of the crystal convert into the global frame. The output layer is initialised to zero, so that training starts from the segment centres. The deposited energy is deliberately not an input to the position output, so that the positions must come from the pulse shapes rather than from the energies that also determine $\theta_E$. It enters only a second output, a learned angular uncertainty for each hit~\cite{Kendall2017}.

\subsection{Loss function}\label{sec:loss}

The residual of pair $i$ is normalised by its expected uncertainty, $r_i=\Delta\theta_i/\sigma_i$, with
\begin{equation}
 \sigma_i^2=\sigma_0^2+\sigma_{E,i}^2+\sigma_{G,i}^2+s_{i1}^2+s_{i2}^2 .
 \label{eq:sigma}
\end{equation}
The constant $\sigma_0=2\dg$ accounts for the spread of the residual that remains even with the packed positions. The term $\sigma_{E,i}$ propagates an energy uncertainty of 1.2~keV per hit, a standard deviation somewhat larger than the simulated resolution, through Eq.~\eqref{eq:compton}. The learned per-hit terms $s_{i1}$ and $s_{i2}$ remain small: for the final network they are below $0.025\dg$ for 99\% of the hits of the validation pairs. Pairs that have no physical angle in the order in use, which step 4 tolerates, are given $\sigma_{E,i}=10^{3}$~rad, so that their $r_i$ is practically zero. The geometrical term approximates the effect of a position uncertainty of 3~mm per hit on the incident and scattered directions,
\begin{equation}
 \sigma_{G,i}=3~\mathrm{mm}\,\sqrt{\ell_{{\rm in},i}^{-2}+2\,\ell_{{\rm sc},i}^{-2}} ,
 \label{eq:lever}
\end{equation}
where $\ell_{{\rm in},i}$ is the distance from the source to the first hit and $\ell_{{\rm sc},i}$ the distance between the hits; the factor 2 arises because the scattered direction has an uncertain position at both ends. Both lever arms are computed from the segment centres, with a lower bound of 8~mm, so that the network cannot reduce its loss by moving the hits closer together.

The first training stage uses the negative log-likelihood of a Cauchy distribution, which limits the influence of pairs that violate the two-interaction hypothesis, with a weak penalty on the displacements,
\begin{equation}
\begin{split}
 \mathcal L={}&\frac{1}{N}\sum_{i=1}^{N}\left[\ln\!\left(1+r_i^2\right)+\ln\sigma_i\right]\\
 &+\frac{\lambda}{N a^2}\sum_{i=1}^{N}\left(\|\boldsymbol\delta_{i1}\|^2+\|\boldsymbol\delta_{i2}\|^2\right),
\end{split}
 \label{eq:loss}
\end{equation}
where $N$ is the number of pairs in a batch, $\boldsymbol\delta_{i1}$ and $\boldsymbol\delta_{i2}$ are the displacements of the two hits and angles are in radians. The term $\ln\sigma_i$ prevents the network from reducing all $|r_i|$ by increasing its learned uncertainties. The penalty, with $\lambda=0.02$ and $a=12$~mm, pulls each hit weakly towards its segment centre and thus damps displacements along directions that the constraint leaves undetermined (Sect.~\ref{sec:degeneracy}). In the later stages, the Cauchy term is replaced by a trimmed mean~\cite{Rousseeuw1984}: in each batch, the 20\% of pairs with the largest $|r_i|$ are discarded, and
\begin{equation}
\begin{split}
 \mathcal L_{\rm trim}={}&\frac{1}{|K|}\sum_{i\in K}|r_i|\\
 &+\frac{\lambda}{N a^2}\sum_{i=1}^{N}\left(\|\boldsymbol\delta_{i1}\|^2+\|\boldsymbol\delta_{i2}\|^2\right),
\end{split}
 \label{eq:trimmed}
\end{equation}
where $K$ is the set of retained pairs. Unlike the Cauchy loss, whose gradient decreases but never vanishes for large residuals, the trimmed loss ignores the largest residuals and weights all other pairs equally. Since it has no $\ln\sigma_i$ term, $\sigma_i$ then contains only the fixed terms $\sigma_0$, $\sigma_{E,i}$ and $\sigma_{G,i}$.

\subsection{Training procedure}\label{sec:training}

A random subset of $1.85\times10^{4}$ pairs of the initial sample is set aside as \emph{validation pairs}, never used to compute gradients; their $3.70\times10^{4}$ hits are the \emph{held-out hits}. The remaining $2.19\times10^{6}$ pairs of the initial sample form the first \emph{training pool}. The network is optimised with AdamW~\cite{Loshchilov2019} in batches of 2000 pairs; an epoch is one pass over a random subset of $5\times10^{5}$ pairs of the training pool, redrawn every five epochs. In each stage the learning rate rises during a short warm-up and then follows a cosine decay~\cite{Loshchilov2017} (Table~\ref{tab:parameters}). The weights are averaged with an exponential moving average~\cite{Polyak1992}, and the averaged weights after the last epoch of a stage are its result and the start of the next stage. Training has four stages:
\begin{enumerate}
\item \emph{initial training}: stored order, first training pool, Cauchy loss, 60 epochs;
\item \emph{selection refinement}: stored order, training pool selected in step 7 with the stage-1 network ($1.41\times10^{6}$ pairs), trimmed loss, 60 epochs;
\item \emph{reordering}: order renewed and training pool selected with the stage-2 network ($1.45\times10^{6}$ pairs), trimmed loss, 60 epochs;
\item \emph{extended fine-tuning}: as stage 3, with order and selection renewed with the stage-3 network ($1.46\times10^{6}$ pairs), 200 epochs.
\end{enumerate}
The validation pairs are removed from every training pool. The network after stage 4 is the \emph{self-calibrated network}. To renew the order at the start of stages 3 and 4, both orders of every pair of the base sample are evaluated with the network of the previous stage, which serves both since it reconstructs each hit independently. If only one order gives a physical angle $\theta_E$, it is taken; otherwise the order with the smaller $\dd$ is taken.

In all stages, Gaussian noise with a standard deviation of 2\% of the normalised pulse amplitude, i.e.\ 4~keV for a 200~keV deposit and 28~keV for a 1400~keV deposit, is added to the input pulses as a regulariser. This \emph{training-only noise} is not applied when a network is evaluated, and the self-calibration needs it: in a test without it, stopped after 88 epochs of stage 1 of the preliminary training described below, the mean position error of the hits of the preliminary validation pairs was 3.09~mm, compared with 2.66~mm with it, although the median $\dd$ was the same, $3.33\dg$, in both runs.

Some results come from an earlier \emph{preliminary training} with the same network and detector response. It had no energy threshold (step 5), so that steps 6 and 7 were applied to all $3.53\times10^{6}$ stored pairs and its $2\times10^{4}$ \emph{preliminary validation pairs} include pairs with a hit below 200~keV. It used the Cauchy loss in all stages and only three stages, of 95, 40 and 40 epochs, and the result of each stage was the epoch with the smallest median $\dd$ of its validation pairs. Its stage-3 result is the \emph{preliminary network}.

The calculations ran on one NVIDIA Quadro GV100 GPU with TensorFlow and Keras~\cite{Abadi2016,Chollet2015}. With an XLA-compiled training step and the detector response of the next pulses computed in parallel, an epoch takes 7~s instead of 17~s, and the complete self-calibration, including the renewals of selection and order, takes 52~min. The trained network reconstructs $1.4\times10^{6}$ hits per second on the GPU ($4\times10^{6}$ in half precision) and $7.5\times10^{4}$ on a 10-core CPU.

\subsection{Supervised reference and evaluation}\label{sec:supervised}\label{sec:evaluation}

To separate the limitations of the constraint from those of the network and the pulses, a \emph{supervised network} with the same architecture and inputs is trained with the packed positions as labels, using a Huber loss~\cite{Huber1964} on the three-dimensional distance (quadratic below 3~mm). It is trained in a single stage of 200 epochs, with a peak learning rate of $10^{-3}$, on the training pool and order of stage 3, with the same detector response but without the training-only noise. Its mean position error on the held-out hits, 1.07~mm, has nearly converged; a 40-epoch training gives 1.35~mm, so that the gap to the self-calibration depends on the length of the supervised training.

Unless stated otherwise, position errors are evaluated for the held-out hits with superposed pulses. Since new noise is drawn whenever a pulse is loaded, all networks are evaluated on the same noisy pulses, the \emph{common noise realisation}; values monitored during training (Fig.~\ref{fig:training}, Table~\ref{tab:stages}) used other realisations and differ slightly. Confidence intervals at 95\% are obtained by bootstrap resampling of the evaluated pairs, for a difference between two networks with the same resampled pairs for both (\emph{paired bootstrap}). Because the validation pairs were drawn by pair rather than by event, 0.5\% of the held-out hits also belong to a training pair with another hit of the same event; since the networks do not overfit (Sect.~\ref{sec:accuracy}), this overlap does not affect the results.

\section{Results}\label{sec:results}

\subsection{Training stages}\label{sec:stages}

\begin{figure*}[t]
\centering
\includegraphics[width=\textwidth]{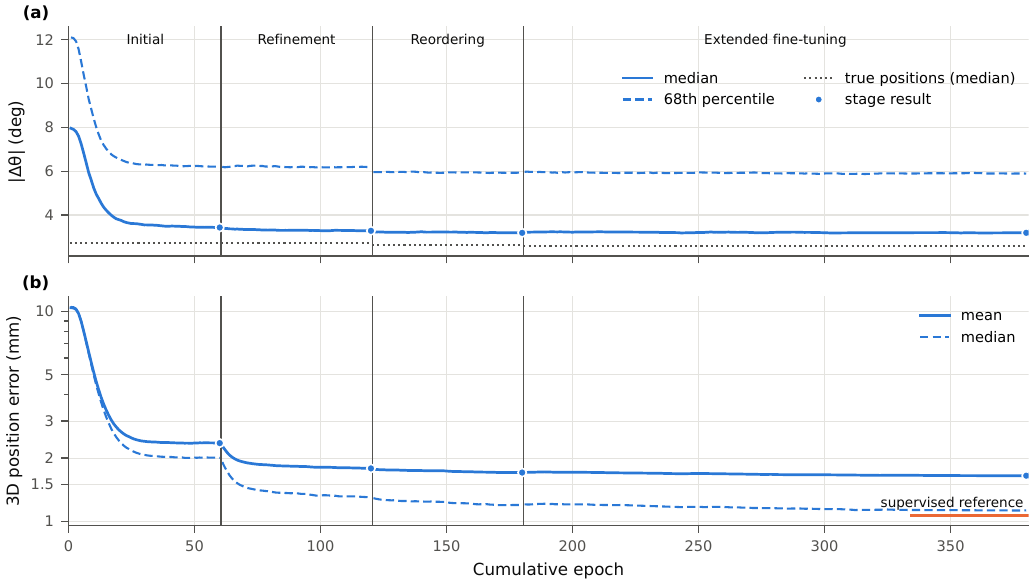}
\caption{Validation metrics during the four stages of the self-calibration, versus the epoch counted from the start of stage~1. (a) Median and 68th percentile of $\dd$ for the validation pairs, computed with the reconstructed positions. Dotted line (true positions): median $\dd$ computed with the packed positions for the same pairs and order; it drops at the start of stage~3, when the order is renewed. (b) Mean and median three-dimensional position error of the held-out hits (logarithmic scale). The horizontal bar (supervised reference) marks the final mean error of the supervised network. Circles (stage result) mark the values after the last epoch of each stage, listed in Table~\ref{tab:stages}}
\label{fig:training}
\end{figure*}

Figure~\ref{fig:training} and Table~\ref{tab:stages} follow the training on the validation pairs. Although only the angular residual enters the loss, the residual and the position error decrease together: starting from the segment centres, with a mean position error of 10.41~mm and a median $\dd$ of $7.95\dg$, the four stages reduce the mean position error of the held-out hits to 2.36, 1.79, 1.71 and 1.65~mm. Stage 4 still gains 0.06~mm, of which only 0.002~mm in its last 20 epochs.

The result of each stage is the network after its last epoch rather than a checkpoint chosen by the validation residual, the only criterion available for measured data. In stages 2--4, the median $\dd$ fluctuates by about $0.01\dg$ from epoch to epoch while the position error still decreases, so that it does not identify the most accurate epoch. In an earlier run with the same recipe but 40-epoch stages, each started from the epoch of the previous stage with the smallest median $\dd$, the stage-3 epoch with the smallest median $\dd$ had a mean position error 0.024~mm larger than the last epoch.

The error is robust to the details of the training. We compared three variants, trained up to stage 3, with the stage-3 network on the common noise realisation, on which the latter has a mean position error of 1.713~mm. A second random seed gives 1.709~mm, and the coordinate residuals of the two networks (Sect.~\ref{sec:accuracy}) are correlated with a coefficient of 0.94 along each axis. Discarding 30\% instead of 20\% of the pairs in the trimmed loss changes the mean error by less than 0.003~mm. Using the Cauchy loss in all three stages, with another seed, gives 1.790~mm, worse by 0.077~mm (paired 95\% interval 0.072--0.083~mm). Since the two seeds differ by only 0.004~mm and give strongly correlated residuals, the error is dominated by components common to all networks trained with the constraint rather than by run-to-run fluctuations.

Renewing the order at the start of stage 3 changes it for 2.9\% of the pairs of the base sample. When both orders give a physical angle, their values of $\dd$ differ by $65\dg$ in the median and by more than $21\dg$ for 90\% of these pairs, so that the choice is unambiguous for most of them. Its effect on the positions is small. In the preliminary training, stage 3 was run twice from the same stage-2 network, with the renewed and with the stored order; after the last epoch, the mean position errors of the hits of the preliminary validation pairs were 2.050 and 2.054~mm (difference 0.004~mm, 95\% interval 0.002--0.006~mm), whereas both runs improved by 0.11~mm during that stage. The gain of stages 3 and 4 thus comes mainly from the renewed selection and the continued training.

\begin{table*}[t]
\caption{Result of each training stage on the validation pairs: $\dd$ of the pairs and three-dimensional position error of their hits, computed with the positions named in the first column. Order: stored, or renewed at the start of the stage given in parentheses. The values were monitored during training, with their own noise draws, and differ slightly from those on the common noise realisation (Table~\ref{tab:heldout}), e.g.\ 1.649~mm here and 1.653~mm there after stage~4. The segment-centre median $\dd$ is smaller than the $10.9\dg$ of the base sample because the validation pairs belong to the initial sample}
\label{tab:stages}
\centering\small
\begin{tabular}{@{}llccccc@{}}
\toprule
 & & & \multicolumn{2}{c}{$\dd$ (\dg)} & \multicolumn{2}{c}{Position error (mm)}\\
\cmidrule(lr){4-5}\cmidrule(l){6-7}
Positions from & Order & Epochs & median & 68th pct. & mean & median\\
\midrule
Segment centres & stored & -- & 7.95 & 12.11 & 10.41 & 10.49\\
Stage~1: initial training & stored & 60 & 3.431 & 6.221 & 2.357 & 2.004\\
Stage~2: selection refinement & stored & 60 & 3.280 & 6.180 & 1.786 & 1.299\\
Stage~3: reordering & renewed (stage~3) & 60 & 3.186 & 5.931 & 1.706 & 1.195\\
Stage~4: extended fine-tuning & renewed (stage~4) & 200 & 3.188 & 5.887 & 1.649 & 1.127\\
Supervised network & renewed (stage~3) & 200 & 2.962 & 5.663 & 1.069 & 0.721\\
\bottomrule
\end{tabular}
\end{table*}
\subsection{Position accuracy on held-out hits}\label{sec:accuracy}

\begin{figure*}[t]
\centering
\includegraphics[width=\textwidth]{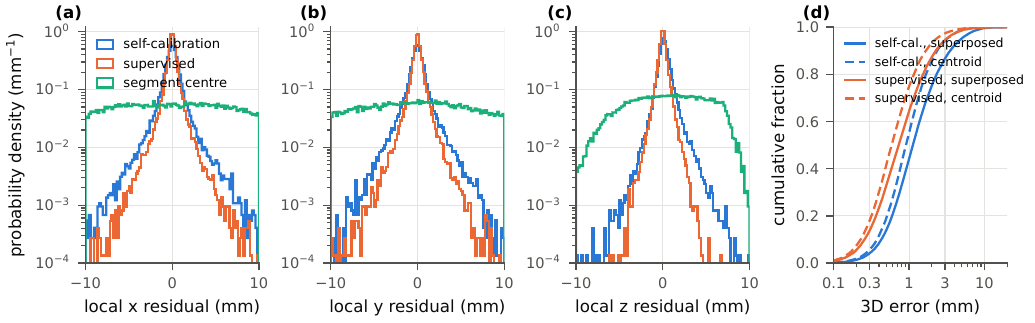}
\caption{Coordinate residuals and position errors of the held-out hits. (a)--(c) Probability density of the coordinate residual along the local $x$, $y$ and $z$ axes, with superposed pulses, for both networks and the segment centres; logarithmic vertical scale, to show the tails. (d) Cumulative fraction of the held-out hits with a three-dimensional position error below a given value, for both networks with superposed and with centroid pulses; logarithmic error axis}
\label{fig:residuals}
\end{figure*}

Table~\ref{tab:heldout} gives the position errors of the held-out hits, and Fig.~\ref{fig:residuals} their distributions. Along each local axis, we call the reconstructed minus the packed coordinate the \emph{coordinate residual}, to distinguish it from the angular residual $\Delta\theta$; Table~\ref{tab:heldout} gives the FWHM of its distribution, which describes the central peak, and its RMS, which includes the tails.

\begin{table*}[t]
\caption{Position errors of the $3.70\times10^{4}$ held-out hits, on the common noise realisation (Sect.~\ref{sec:evaluation}). 3D error: distance between the reconstructed and the packed position. FWHM and RMS: widths of the distribution of the coordinate residual along each local axis ($z$: crystal axis). The RMS along $y$, not listed, is close to that along $x$. The segment centres do not depend on the pulses. Uncertainties are half-widths of 95\% bootstrap intervals}
\label{tab:heldout}
\centering\small
\begin{tabular}{@{}llcccccccc@{}}
\toprule
 & & \multicolumn{3}{c}{3D error (mm)} & \multicolumn{3}{c}{FWHM (mm)} & \multicolumn{2}{c}{RMS (mm)}\\
\cmidrule(lr){3-5}\cmidrule(lr){6-8}\cmidrule(l){9-10}
Pulses & Positions & mean & median & 90th pct. & $x$ & $y$ & $z$ & $x$ & $z$\\
\midrule
-- & Segment centres & 10.41 & 10.49 & 14.97 & 20.5 & 20.1 & 12.9 & 7.13 & 4.31\\
\midrule
\multirow{2}{*}{Superposed} & Self-calibrated & $1.653\pm0.016$ & $1.125\pm0.011$ & 3.51 & 1.27 & 1.23 & 0.98 & 1.48 & 1.00\\
 & Supervised & $1.075\pm0.011$ & $0.724\pm0.007$ & 2.26 & 0.74 & 0.74 & 0.68 & 0.99 & 0.63\\
\midrule
\multirow{2}{*}{Centroid} & Self-calibrated & $1.213\pm0.011$ & $0.907\pm0.008$ & 2.36 & 1.15 & 1.11 & 0.88 & 1.04 & 0.62\\
 & Supervised & $0.839\pm0.008$ & $0.587\pm0.006$ & 1.68 & 0.68 & 0.68 & 0.63 & 0.77 & 0.48\\
\bottomrule
\end{tabular}
\end{table*}

With superposed pulses, the self-calibrated network has a mean position error of 1.653~mm (95\% interval 1.637--1.669~mm) and a median of 1.125~mm, and 90\% of the held-out hits have an error below 3.51~mm. Its coordinate residuals have a narrow central peak, with a FWHM of 1.27, 1.23 and 0.98~mm in $x$, $y$ and $z$, and broader tails (Fig.~\ref{fig:residuals}a--c). The supervised network has a mean error of 1.075~mm, 0.58~mm smaller (paired 95\% interval 0.57--0.59~mm), and a FWHM of 0.74, 0.74 and 0.68~mm in $x$, $y$ and $z$. Both are far more accurate than the segment centres, whose mean error is 10.41~mm and whose coordinate residuals have no central peak.

Both networks are more accurate with centroid pulses, although neither was trained on them: the mean error falls from 1.653 to 1.213~mm for the self-calibrated network and from 1.075 to 0.839~mm for the supervised network. Centroid pulses lack two features of the superposed pulses, the superposition of several basis pulses in a multi-interaction hit and, in a same-crystal pair, the signal of the partner hit. For the self-calibrated network, the gain of 0.44~mm comes in about equal parts from two groups of hits. The hits of same-crystal pairs, 22\% of the held-out hits, are affected by both features, and their mean error falls from 2.13~mm with superposed pulses to 1.10~mm with centroid pulses. For the other 78\% of the held-out hits, only multi-interaction hits differ between the two representations, and their mean error falls from 1.52 to 1.24~mm.

The networks do not overfit. For $2\times10^{5}$ pairs drawn at random from the first training pool (the \emph{training-pool subset}), the mean error of both networks, with both pulse representations, differs from that of the held-out hits by less than 0.6\%. A further $5.18\times10^{5}$ pairs, 18.6\% of the base sample, failed step 6 and all three selections of step 7 and were therefore never used for training. These rejected pairs are harder to reconstruct: in a random subset of $2\times10^{4}$ of them, 64\% of the hits are multi-interaction hits, compared with 45\% of the held-out hits, and the mean error with superposed pulses is 1.87~mm for the self-calibrated network and 1.25~mm for the supervised network.

\subsection{Dependence on energy, crystal and position}\label{sec:spatial}

\begin{figure}[t]
\centering
\includegraphics[width=\columnwidth]{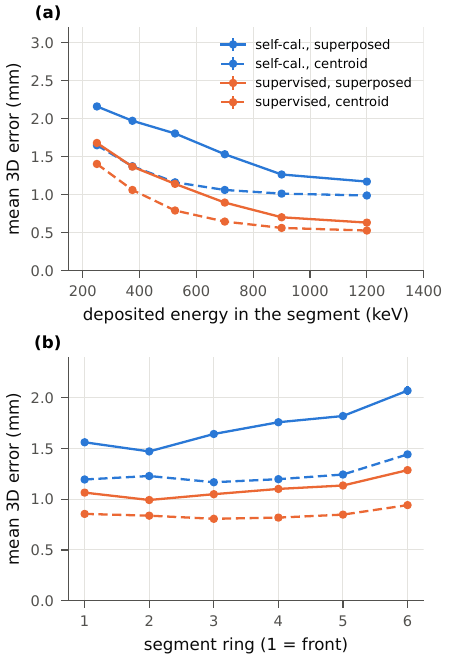}
\caption{Mean position error of the held-out hits versus (a) the energy deposited in the segment and (b) the segment ring, for both networks, with superposed and with centroid pulses. Error bars (standard error of the mean) are mostly smaller than the markers}
\label{fig:energy}
\end{figure}

The mean error decreases with the energy deposited in the segment (Fig.~\ref{fig:energy}a), since the signal amplitude grows with the energy whereas the noise does not. For the self-calibrated network it falls from 2.16~mm at 200--300~keV to 1.17~mm above 1~MeV, and for the supervised network from 1.68 to 0.63~mm. Hits of same-crystal pairs contribute most at low energy: at 200--300~keV, the self-calibrated network has a mean error of 3.16~mm for them and 1.91~mm for the other hits, because their traces, in particular the shared core trace, are dominated by the signal of a partner hit of much larger energy.

The accuracy is uniform across crystals. The mean errors of the self-calibrated network for the three crystal shapes agree within 3\%, and those of the 51 crystals, 1.47--1.91~mm, scatter by 0.08~mm, as expected from their statistical uncertainties of 0.03--0.18~mm ($\chi^2=40$ for 50 degrees of freedom).

\begin{figure*}[t]
\centering
\includegraphics[width=\textwidth]{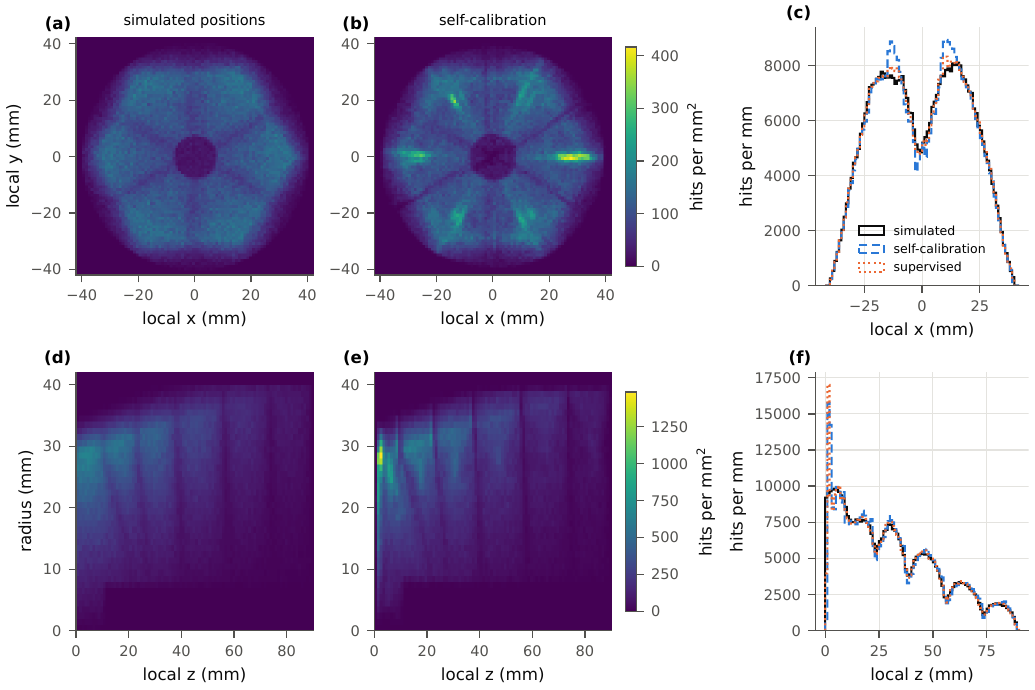}
\caption{Simulated and reconstructed positions of $4.37\times10^{5}$ hits in the local crystal frame, all crystals combined: the held-out hits and the hits of the training-pool subset (Sect.~\ref{sec:accuracy}). Reconstructed positions use superposed pulses. Upper row: transverse plane. Lower row: radius versus depth $z$. (a), (d) Simulated positions. (b), (e) Positions reconstructed by the self-calibrated network. The colour gives the number of hits per mm$^2$, on one scale for the two maps of each row. (c), (f) Distributions, in hits per mm, of the local $x$ coordinate and of the depth, for the simulated positions and for both networks}
\label{fig:positions}
\end{figure*}

From the second ring on, the error increases towards the back of the crystal (Fig.~\ref{fig:energy}b): for the self-calibrated network it is 1.56~mm in the first ring, 1.47~mm in the second and 2.07~mm in the sixth, and for the supervised network it rises from 0.99~mm in the second ring to 1.29~mm in the sixth. The error also depends on the distance of the hit from the crystal axis. To study this and the position distributions with more statistics, we add the hits of the training-pool subset to the held-out hits, $4.37\times10^{5}$ hits in total. The self-calibrated network has its smallest mean error, 1.1~mm, at radii of 10--15~mm and its largest, 1.9--2.0~mm, at 20--30~mm; the supervised network follows the same trend at 0.7 and 1.2--1.3~mm. Neither network has larger errors near the sector boundaries.

\begin{table*}[t]
\caption{Absolute angular residual $\dd$, in degrees, from simulated quantities, for pairs in the stored order unless stated otherwise. Upper part: true-position residual, i.e.\ with packed positions (Sect.~\ref{sec:truth}). Lower part: decomposition for the initial sample (Sect.~\ref{sec:floor}), each line removing one more effect. Line~5 comes from a separate Geant4 run}
\label{tab:truth}
\centering\small
\begin{tabular}{@{}lrccc@{}}
\toprule
Sample and simulated quantities & Pairs & median & 68th pct. & 90th pct.\\
\midrule
Base sample & $2.79\times10^{6}$ & 4.10 & 10.42 & 44.42\\
Initial sample (after step~6) & $2.21\times10^{6}$ & 2.71 & 5.56 & 16.04\\
Validation pairs, stored order & $1.85\times10^{4}$ & 2.71 & 5.54 & 16.45\\
Validation pairs, order renewed before stage~4 & $1.85\times10^{4}$ & 2.60 & 5.29 & 15.24\\
\midrule
(1) Initial sample, packed positions and smeared energies & $2.21\times10^{6}$ & 2.71 & 5.56 & 16.04\\
(2) \quad without energy smearing & $2.21\times10^{6}$ & 2.71 & 5.55 & 16.05\\
(3) \quad and first interaction in each segment & $2.21\times10^{6}$ & 1.57 & 4.16 & 16.12\\
(4) \quad and only pairs with one interaction in the first segment & $1.76\times10^{6}$ & 0.97 & 2.15 & 7.84\\
(5) Doppler broadening only, scattering angles of line~4 & $1.76\times10^{6}$ & 0.82 & 1.76 & 5.84\\
\bottomrule
\end{tabular}
\end{table*}

The self-calibrated positions (Fig.~\ref{fig:positions}b, e) reproduce the hexagonal outline, the ring boundaries and the decrease of the hit density with depth of the simulated positions (Fig.~\ref{fig:positions}a, d), with two distortions. First, the self-calibrated network concentrates hits in a few small regions of each sector at radii of 20--30~mm, where its errors are largest, and places too few along the sector boundaries. In 1~mm$\times$1~mm transverse bins with at least 20 simulated positions, 7.3\% of the self-calibrated positions fall in bins with more than 1.5 times the simulated density, compared with 2.0\% of the supervised positions. Second, both networks reconstruct the hits closest to the front face too deep: the hits simulated within 2~mm of the face, at a mean depth of 1.0~mm, are reconstructed at 1.9~mm on average by the self-calibrated network and at 1.4~mm by the supervised network, consistent with the mean displacement of the self-calibrated positions towards the back of the crystal (Sect.~\ref{sec:degeneracy}).

\begin{figure*}[t]
\centering
\includegraphics[width=\textwidth]{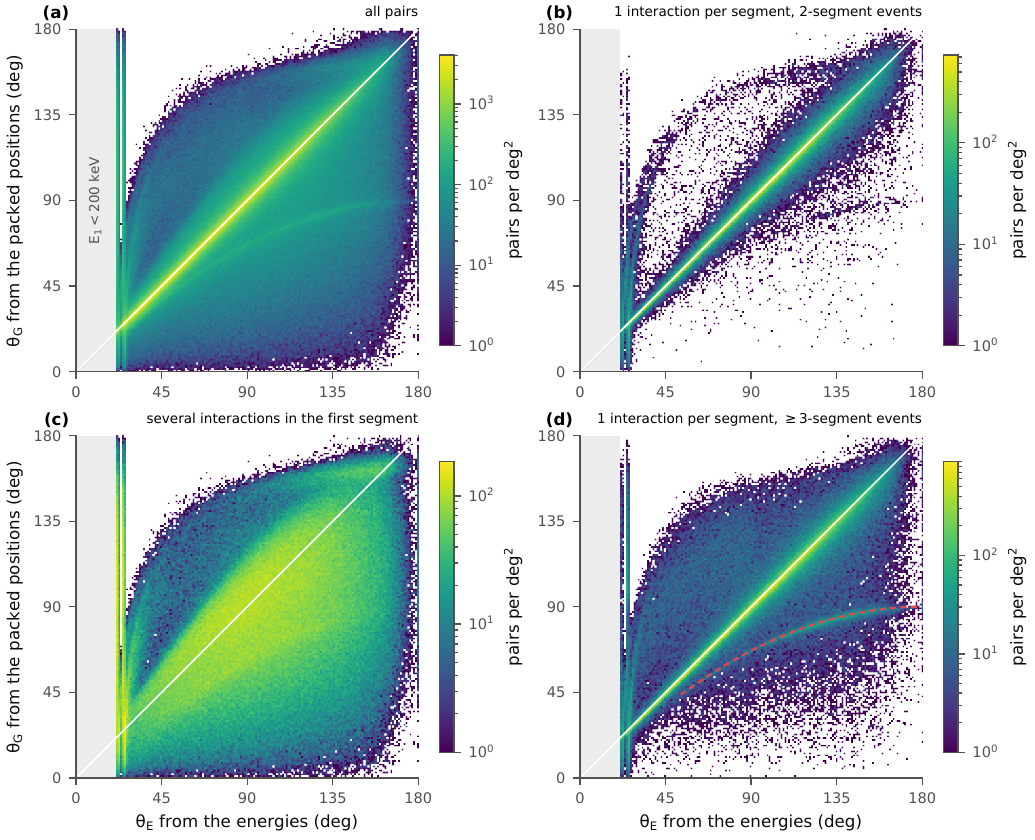}
\caption{Scattering angle $\theta_G$ from the packed positions versus $\theta_E$ from the energies, for the base sample in the stored order: (a) all pairs; (b)--(d) the populations marked (b)--(d) in Table~\ref{tab:populations}. Dashed curve in (d): relation expected for wrong-line pairs. White line: diagonal $\theta_G=\theta_E$. The grey region, $\theta_E<21.2\dg$, is empty because step~5 requires $E_1\ge200$~keV}
\label{fig:band}
\end{figure*}

\subsection{Angular residuals and sample composition}\label{sec:truth}

The true-position residual (Sect.~\ref{sec:constraint}) is the reference against which the networks can be judged. In the stored order, its median is $4.10\dg$ for the base sample and $2.71\dg$ for the initial sample (Table~\ref{tab:truth}). For the validation pairs in the order renewed before stage 4, in which the networks are evaluated, it is $2.60\dg$. On the same pairs and pulses, the median $\dd$ is $3.18\dg$ with the self-calibrated network, 22\% above this reference, and $3.01\dg$ with the much more accurate supervised network; the 68th percentiles are $5.94\dg$ and $5.65\dg$.

\begin{table}[t]
\caption{Populations of the base sample, defined in the stored order. Letters: panels of Fig.~\ref{fig:band}. Share: fraction of the base sample. Kept: fraction of the population passing step~6 or the step-7 selection made before stage~4, each applied to the whole base sample. The four middle populations partition the base sample, and the last three overlap with them. Wrong line: pairs on the 1173~keV line with $\theta_E>60\dg$ and $\theta_G$ within $2\dg$ of the dashed curve of Fig.~\ref{fig:band}d, minus the background beside the curve}
\label{tab:populations}
\centering\small
\setlength{\tabcolsep}{4pt}
\begin{tabular}{@{}>{\raggedright\arraybackslash}p{0.5\columnwidth}rrr@{}}
\toprule
 & Share & \multicolumn{2}{c}{Kept (\%)}\\
\cmidrule(l){3-4}
Population & (\%) & step~6 & step~7\\
\midrule
All pairs of the base sample (a) & 100 & 79.1 & 52.4\\
\midrule
One interaction in each segment, two fired segments (b) & 8.4 & 92.5 & 76.8\\
One interaction in each segment, three or more fired segments (d) & 13.2 & 80.6 & 61.3\\
One interaction in the first segment, several in the second & 49.6 & 89.7 & 67.1\\
Several interactions in the first segment (c) & 28.7 & 56.3 & 15.6\\
\midrule
Reversed order & 3.1 & 70.4 & 43.4\\
Wrong line & 0.5 & 43.7 & 0.4\\
Imposed order & 3.3 & 25.4 & 14.5\\
\bottomrule
\end{tabular}
\end{table}

Figure~\ref{fig:band} shows $\theta_G$, computed with the packed positions, against $\theta_E$ for the base sample in the stored order, and Table~\ref{tab:populations} lists the populations that the recorded interactions distinguish. Pairs with a single interaction in each segment lie on the diagonal, both in events with two fired segments (Fig.~\ref{fig:band}b; median true-position residual $1.2\dg$) and in events with more (Fig.~\ref{fig:band}d), and pairs with one interaction in the first segment and several in the second, half of the base sample, stay close to it (median $2.4\dg$). The pairs with several interactions in the first segment, 28.7\% of the base sample, form a broad cloud (Fig.~\ref{fig:band}c, median $21\dg$), because their $E_1$ includes deposits made after the first scattering. Three further structures leave the diagonal. A thin curve above it is formed by the pairs whose stored order is reversed, for which $\theta_G$ is measured at the absorption point. A narrow band below it, levelling off at $\theta_G\approx90\dg$, is formed by wrong-line pairs: the first two interactions of a 1332~keV photon that left about 159~keV in a third segment, accepted on the 1173~keV line, so that $\theta_E$ is computed for the wrong photon energy (dashed curve in Fig.~\ref{fig:band}d); 94\% of them come from events with three or more fired segments. Finally, vertical stripes at $\theta_E\approx21\dg$--$22\dg$ and $24\dg$--$25\dg$ contain the imposed-order pairs, whose $E_1$ lies between 200~keV and the backscatter energy, so that only the stored order is allowed; for 68\% of those with a true-position residual above $25\dg$, the photon in fact scattered several times in the segment stored second and was absorbed in the one stored first.

Except for the imposed-order pairs, these populations cannot be identified in a measurement, but steps 6 and 7 remove most of the inconsistent pairs (Table~\ref{tab:populations}). The step-7 selection made before stage 4 keeps 52.4\% of the base sample, but 76.8\% of the pairs with one interaction per segment from two-segment events, 15.6\% of the pairs with several interactions in the first segment and only 0.4\% of the wrong-line pairs.

\subsection{Origin of the true-position residual}\label{sec:floor}

Since the simulation records every interaction, the contributions to the median true-position residual of the initial sample, $2.71\dg$, can be removed one at a time (lower part of Table~\ref{tab:truth}). Without the energy smearing, the median changes by less than $0.01\dg$: the angular uncertainty from the energy resolution, with a median of $0.11\dg$, adds in quadrature to a much broader distribution. Replacing the packed positions by the first interaction in each segment lowers the median to $1.57\dg$, entirely through the second hit, because in 69\% of the pairs the scattered photon interacts several times in the second segment. The pairs with several interactions in the first segment, 20\% of the initial sample, have a median residual of $11\dg$--$12\dg$ whatever positions are used, because $\theta_E$ does not describe their first scattering; without them, the median falls to $0.97\dg$ and the 90th percentile from $16.12\dg$ to $7.84\dg$.

Doppler broadening sets the lower limit. In a separate Geant4 run with the same physics, the difference between the deflection angle of the photon and the free-electron angle from its energies has a median of $0.54\dg$ for the first scattering of the 1173 and 1332~keV photons, growing from $0.23\dg$ below $30\dg$ to $2.4\dg$ above $120\dg$. For the scattering angles of the pairs with one interaction in the first segment (line~4 of Table~\ref{tab:truth}), the median is $0.82\dg$ (line~5). The remaining difference to $0.97\dg$ comes from unrecorded deflections between the two segments (Sect.~\ref{sec:packing}), from the reversed-order pairs (2.8\% of the initial sample) and from wrong-line and imposed-order pairs. The description of each segment by a single position and energy is thus the dominant contribution to the true-position residual.

\section{Discussion}\label{sec:discussion}

\subsection{Compton constraint}\label{sec:degeneracy}

Part of the 0.58~mm gap between the self-calibrated and the supervised network is systematic. The self-calibrated positions of the held-out hits are displaced from the packed positions by $+0.22$~mm on average along the crystal axis, towards the back, growing from $+0.14$~mm in the first ring to $+0.40$~mm in the last, and by 0.15~mm radially inwards, whereas the supervised network has no mean displacement above 0.03~mm. The constraint is insensitive to such distortions. Shifting the packed positions of the validation pairs by 1~mm along the crystal axes changes their median $\dd$ (in the stored order, $2.71\dg$) by less than $0.01\dg$, whereas the same shift along the local $x$ axes increases it by $0.16\dg$--$0.19\dg$. The reason is geometric: all crystal axes point at the centre of the array, so an axial shift moves the hits of neighbouring crystals apart nearly along the line that joins them. It rotates the scattered direction by a median of only $0.13\dg$ and the incident direction by $0.26\dg$, compared with $0.91\dg$ and $0.39\dg$ for a shift along $x$. In addition, a uniform scaling of all distances from the source leaves every $\theta_G$ unchanged, so that position maps that locally resemble such a scaling are only weakly penalised. Additional source positions, which change the incident directions differently for each crystal, should constrain the axial direction, and a depth reference such as the front face could fix a remaining offset.

The same comparison shows that the angular residual alone cannot certify the position accuracy. The mean position error of the supervised network is 35\% smaller than that of the self-calibrated network, yet its median $\dd$ on the validation pairs is only $0.17\dg$ lower (Sect.~\ref{sec:truth}). For measured data, the residual remains a diagnostic of the convergence and of the selection, to be complemented by independent checks such as collimated sources or additional source positions.

\subsection{Multiple interactions in a segment}\label{sec:multiple}

The packed position of a multi-interaction second hit differs from the first interaction of the scattered photon in that segment, which defines the scattered direction; this accounts for most of the difference between the true-position residual and the Doppler limit (Sect.~\ref{sec:floor}). The pairs with a multi-interaction first hit violate the two-interaction hypothesis, and most of them fail step 7. The constraint does not pull the reconstructed second hits towards their first interaction. For the multi-interaction second hits of validation pairs in different crystals, the first interaction lies on average 3.7~mm from the packed position, but the self-calibrated positions are displaced towards it by 0.003~mm on average: the single position per hit adds noise to the constraint rather than a bias. In the position estimate itself, the superposed pulses of multi-interaction hits and the partner signal in same-crystal pairs cost accuracy (Sect.~\ref{sec:accuracy}). Reconstructing the hits of a crystal jointly, with the multiple-interaction PSA of AGATA~\cite{Boston2023}, and including all fired segments of the event in the pulse synthesis are natural extensions.

\subsection{Small scattering angles}\label{sec:lowenergy}

Hits below 200~keV, which are always first hits with small scattering angles, are excluded because of a bias found in the preliminary training. To study it, the preliminary network and a \emph{preliminary supervised network}, trained on the same pairs with the packed positions as labels, are evaluated on the preliminary validation pairs, which include such hits. The hits below 100~keV, 1.7\% of the hits of these pairs, with a median $\theta_E$ of $11\dg$, had a mean position error of 13.1~mm with the preliminary network, larger even than with the segment centres, compared with 5.8~mm with the preliminary supervised network. The error was directed towards the source along the incident direction: the mean shift was 17.7~mm below 50~keV and 7.3~mm at 50--100~keV, falling to 1.5~mm at 100--150~keV and 0.2~mm at 150--200~keV, whereas the transverse RMS errors below 100~keV, 4--6~mm, were similar for both networks.

The shift compensates a bias of the geometrical angle. At small angles, $\theta_G$ is roughly the distance of the second hit from the line through the source and the first hit, divided by the distance between the hits. A transverse error of the first position adds in quadrature to that distance and so increases $\theta_G$ on average, whereas moving the first hit towards the source along the line lengthens the distance between the hits and reduces $\theta_G$ for any position of the second hit. The shift thus lowers the angular residual while degrading the positions: for the preliminary validation pairs with a hit below 100~keV, the median $\dd$ was $3.0\dg$ with the preliminary network but $4.2\dg$ with the more accurate preliminary supervised network. The final networks, trained without such hits, still reconstruct the hits below 100~keV of the preliminary validation pairs poorly: the mean error is 8.1~mm for the self-calibrated network, which still shifts them towards the source by 2.5~mm on average, and 6.9~mm for the supervised network. Including them would require a loss with a position uncertainty for each hit instead of the fixed 3~mm of Eq.~\eqref{eq:lever}.

\subsection{Relation to previous work and limitations}\label{sec:comparison}\label{sec:limitations}

The method shares its principle with Refs.~\cite{Heil2018,Chen2025}, but replaces the optimisation of one position per group of interactions, formed from the simulated positions in Ref.~\cite{Heil2018} and by clustering similar pulses in Ref.~\cite{Chen2025}, by a single continuous mapping from the pulses of each hit to a position. It therefore needs no clustering thresholds or selection of well-populated clusters. The $4\times10^{7}$ decays simulated here correspond to about 7~min of a 100~kBq source, compared with $2\times10^{10}$ simulated 2~MeV photons from three source positions in Ref.~\cite{Chen2025}, although the dependence on the number of events was not studied here. The resolutions are not directly comparable: Ref.~\cite{Chen2025} quotes the resolution of the AGATA PSA with the self-calibrated basis, whereas here the network itself is the position estimator. Unlike networks trained on scanning data~\cite{Abushawish2026}, the self-calibration needs neither a scanning system nor labels, but it relies on the known array geometry and source position.

The detector response is modelled: crystal differences are approximated by a stretch of the time axis and variations of cross-talk, gains and delays, whereas real crystals differ in their impurity profiles and, after irradiation, in charge trapping; the noise is filtered white noise, and neither the extent of the charge clouds nor the signals of other fired segments are included. The accuracies of Table~\ref{tab:heldout} are therefore expected to be optimistic for measured data. Neither the benefit of the crystal identifier nor a crystal-by-crystal fine-tuning was tested. An error in the assumed source position would bias $\theta_G$ of all pairs coherently; its effect, a fit of the source position and a test with another source position remain to be studied. The accuracy was compared only with the supervised network, not with the grid-search PSA of AGATA on the same pulses. Finally, four choices of the procedure (the 200~keV threshold, the trimmed loss, the last-epoch rule and the length of stage 4) were made by comparing errors computed with the simulated positions, which are not available for measured data and which the angular residual alone would not have replaced (Sect.~\ref{sec:degeneracy}). Validation with measured pulses is the next step.

\section{Conclusions}\label{sec:conclusions}

In this simulation study, the Compton scattering relation provided enough information to calibrate the pulse-to-position response of a segmented HPGe tracking array without position labels. For 51 AGATA crystals irradiated by a single \Co\ source, with a detector response that includes noise, preamplifier response, cross-talk, trigger-time jitter and differences between crystals, a network trained only to reconcile the geometrical and energy-derived scattering angles of pairs of hits whose energies add up to a \Co\ line reduces the mean three-dimensional error of held-out hits of at least 200~keV from 10.4~mm at the segment centres to 1.65~mm, with a FWHM of 1.0--1.3~mm in each coordinate. A supervised network with the same architecture and inputs reaches 1.07~mm. The calibration of the 51 crystals takes under an hour on a single GPU, and its training and selection use only information available for measured data.

The main limitations are the weak constraint along the crystal axes, which leaves the self-calibrated positions shifted towards the back of the crystal by 0.14--0.40~mm depending on the ring; the description of each segment by a single position and energy, which is the main reason why the median true-position residual of the initial sample, $2.7\dg$, exceeds the Doppler limit of about $0.8\dg$; and the modelled detector response. At small scattering angles, the constraint pulled first hits below 100~keV towards the source, so that deposits below 200~keV are excluded. Measurements with several source positions, which should constrain the positions along the crystal axes, and validation with measured pulses are the next steps towards an in-situ calibration of AGATA.

\bmhead{Acknowledgements}
The authors thank Johan Ljungvall for calculating and providing the AGATAGeFEM basis in 1~mm grid size.

\bmhead{Funding}
To be completed by the authors.

\bmhead{Data Availability Statement}
This manuscript has no associated data or the data will not be deposited. [Authors' comment: The simulated data sets generated and analysed during the current study are available from the corresponding author on reasonable request.]

\bmhead{Code Availability Statement}
Code/software will be made available on reasonable request. [Authors' comment: The code used to train and evaluate the networks and to produce the figures is available from the corresponding author.]

\bmhead{Competing interests}
The authors declare no competing interests.

\bmhead{Author contributions}
To be completed by the authors.

\begin{appendices}
\providecommand{\theHfigure}{}\providecommand{\theHtable}{}%
\renewcommand{\theHfigure}{app\thesection.\arabic{figure}}\renewcommand{\theHtable}{app\thesection.\arabic{table}}%

\section{Network and training settings}\label{app:parameters}

Table~\ref{tab:parameters} lists the settings of the network, of the loss and of the training stages of the final training. The preliminary training differs as listed in Sect.~\ref{sec:training}.

\begin{table*}[t]
\caption{Settings of the self-calibration. The supervised reference uses the same network and data pipeline, with the differences given in Sect.~\ref{sec:supervised}}
\label{tab:parameters}
\centering\small
\setlength{\tabcolsep}{4pt}
\begin{tabular}{@{}ll@{}}
\toprule
Item & Setting\\
\midrule
Input per hit & 6 channels $\times$ 55 samples (10~ns)\\
Encoder & Conv1D 64/7, 64/5 (stride 2), 128/5, 128/3 (stride 2);\\
 & dense 256 + layer norm., dense 192\\
Additional inputs & segment (embedding 8), crystal shape (embedding 4),\\
 & crystal identifier (embedding 8),\\
 & segment centre (local, scaled by 1/50~mm)\\
Position head & dense 256, 128; output $25~\mathrm{mm}\tanh(\cdot)$, zero-initialised\\
Uncertainty head & from last hidden layer and $\ln E$; $8\dg\times\mathrm{sigmoid}(\cdot)$\\
Activation & SiLU\\
Trainable parameters & $7.12\times10^{5}$\\
\midrule
Loss & stage~1: Cauchy loss, Eq.~\eqref{eq:loss}; stages~2--4: trimmed mean of $|r|$,\\
 & 20\% discarded per batch, Eq.~\eqref{eq:trimmed}; $\sigma_0=2\dg$, $\sigma_E$ from 1.2~keV,\\
 & $\sigma_G$ from 3~mm, lever arms $\ge8$~mm; $\lambda=0.02$, $a=12$~mm\\
Optimiser & AdamW, weight decay $10^{-4}$, clip norm 1\\
Batch / epoch & 2000 pairs; 250 batches; patch $5\times10^{5}$ pairs, redrawn every 5 epochs\\
Learning rate & stage~1: $10^{-3}$, 3-epoch warm-up, cosine (60 epochs);\\
 & stages~2, 3: $3\times10^{-4}$, 1-epoch warm-up, cosine (60 epochs);\\
 & stage~4: $3\times10^{-4}$, 1-epoch warm-up, cosine (200 epochs)\\
Weight averaging & exponential, decay 0.999 per batch\\
Detector response & Sect.~\ref{sec:noise} (3~keV noise, cross-talk, gains, delays,\\
 & time stretch, trigger fit)\\
Training-only noise & Gaussian, 2\% of the normalised amplitude (self-calibration)\\
Selection & both hits $\ge200$~keV (step~5); stage~1: segment-centre $\dd\le25\dg$ (step~6);\\
 & stages~2--4: $\dd\le5\dg$ with reconstructed positions (step~7), renewed before each stage\\
Stage result & averaged weights after the last epoch\\
\bottomrule
\end{tabular}
\end{table*}

\end{appendices}

\bibliography{references}

\end{document}